\documentclass[10pt,aps,prd,twocolumn,floats,floatfix,showpacs,superscriptaddress,nofootinbib]{revtex4-1}
\usepackage[utf8]{inputenc}               		

\usepackage{graphicx,mathtools,amssymb,amsmath,amsthm,amsfonts,epsfig,epsf}
\usepackage[outdir=./]{epstopdf}
\usepackage[dvipsnames]{xcolor}
\usepackage[linktocpage]{hyperref}
\hypersetup{
    colorlinks=true,
    linkcolor=blue,
    filecolor=magenta,      
    urlcolor=BrickRed,
    citecolor=BrickRed,
}
\hypersetup{pdfstartview=}	
\usepackage{tensor}	
\usepackage{csquotes}
\usepackage{mathrsfs}

\usepackage{graphicx}
\usepackage{amsmath}
\usepackage{autobreak}

\newcommand{\dd}{\mathrm{d}}

\newcommand{\gb}{\mathcal{G}}

\newcommand{\meff}{m_{\text{eff}}}
\newcommand{\beq}{\begin{equation}}
\newcommand{\eeq}{\end{equation}}

\begin{document}

\allowdisplaybreaks

\title{Black hole scalarization with Gauss-Bonnet and Ricci scalar couplings}

\author{\textbf{Georgios Antoniou}}
\affiliation{School of Mathematical Sciences, University of Nottingham,
University Park, Nottingham NG7 2RD, United Kingdom}

\author{\textbf{Antoine Lehébel}}
\affiliation{Centro de Astrof\'{\i}sica e Gravita\c c\~ao  - CENTRA,
Departamento de F\'{\i}sica, Instituto Superior T\'ecnico - IST,
Universidade de Lisboa - UL,
Av. Rovisco Pais 1, 1049-001 Lisboa, Portugal}

\author{\textbf{Giulia Ventagli}}
\affiliation{School of Mathematical Sciences, University of Nottingham,
University Park, Nottingham NG7 2RD, United Kingdom}

\author{\textbf{Thomas P.~Sotiriou}}
\affiliation{School of Mathematical Sciences, University of Nottingham,
University Park, Nottingham NG7 2RD, United Kingdom}
\affiliation{School of Physics and Astronomy, University of Nottingham,
University Park, Nottingham NG7 2RD, United Kingdom}

%%%

\begin{abstract}
Spontaneous scalarization is a gravitational phenomenon in which deviations from general relativity arise once a certain threshold in curvature is exceeded, while being entirely absent below that threshold. For black holes, scalarization is known to be triggered by a coupling between a scalar and the Gauss-Bonnet invariant. A coupling with the Ricci scalar, which can trigger scalarization in neutron stars, is instead known to not contribute to the onset of black hole scalarization, and has so far been largely ignored in the literature when studying scalarized black holes. In this paper, we study the combined effect of both these couplings on black hole scalarization. We show that the Ricci coupling plays a significant role in the properties of scalarized solutions and their domain of existence. This work is an important step in the construction of scalarization models that evade binary pulsar constraints and have general relativity as a cosmological late-time attractor, while still predicting deviations from general relativity in black hole observations.
\end{abstract}

\maketitle

%%%

\section{Introduction}

Neutron stars and black holes constitute environments where gravitational effects can be highly nonlinear and curvatures are the strongest we currently have access to. Since 2015, we have a direct observation channel to this regime through the binary mergers observed by LIGO and Virgo collaborations \cite{Abbott:2016blz,TheLIGOScientific:2017qsa,Abbott:2020niy}. The exploration of the deviations from general relativity (GR) or extensions of the standard model with gravitational waves is incredibly promising and still in its dawn (see {\em e.g.}~\cite{Barack:2018yly,Sathyaprakash:2019yqt,Barausse:2020rsu}). Future detectors will greatly improve the sensitivity and scope of detection of putative new fundamental fields. 

New fields that manage to avoid detection at lower curvatures and yet can significantly affect the properties of compact objects deserve particular attention in this context. Scalarization provides a mechanism for that. Originally proposed by Damour and Esposito-Farèse (DEF) \cite{Damour:1992kf} in the context of neutron stars, scalarization is a phase transition phenomenon that unfreezes a scalar degree of freedom above a critical curvature scale. From a perturbative standpoint, DEF scalarization can be understood as follows. The scalar-tensor theory admits as solutions spacetimes that solve Einstein's equation with a constant scalar field. At low curvatures these solutions are stable. However, cranking up the curvature endows the scalar with a tachyonic mass, which destabilizes the spacetime.
The growth of the scalar is ultimately quenched by non-linearities and the outcome is a spacetime with a non-trivial scalar configuration that deviates from the GR solution. 
Pulsar data severely constrains the original DEF model \cite{Freire:2012mg,Antoniadis:2013pzd,Shao:2017gwu} but the constraint can be evaded by the addition of a bare mass \cite{Ramazanoglu:2016kul}.

 In the DEF model, the scalar field couples to the Ricci scalar, $R$, and it is this coupling that controls the effective mass of the scalar field in curved backgrounds. For black hole spacetimes that solve Einstein's equations in vacuum, $R=0$ and the scalar does not acquire an effective mass. Hence, black holes cannot scalarize (unless scalarization is triggered by surrounding matter \cite{Cardoso:2013opa,Cardoso:2013fwa}). Indeed, the model is subject to no-hair theorems \cite{Hawking:1972qk,Sotiriou:2011dz}. More recently however, it has been pointed out that a coupling between the scalar and the Gauss-Bonnet invariant 
%%%%
\begin{equation}
    \gb = R_{\mu\nu\rho\sigma} R^{\mu\nu\rho\sigma}
- 4 R_{\mu\nu} R^{\mu\nu} + R^2
\end{equation}
%%%%
 can lead to scalarization of black holes, as well as neutron stars \cite{Silva:2017uqg,Doneva:2017bvd}. Reference~\cite{Andreou:2019ikc} later identified all possible couplings that can trigger the tachyonic instability associated with  scalarization (one can also consider different field contents or instabilities, {\em e.g.} \cite{Ramazanoglu:2017xbl,Ramazanoglu:2018hwk,Herdeiro:2018wub,Doneva:2018rou}). The scalarization thresholds for a neutron star background were computed in Ref.~\cite{Ventagli:2020rnx} and they are indeed predominantly controlled by the couplings to the Ricci and Gauss-Bonnet scalars  (and potentially a bare mass). For black holes, the analysis of Ref.~\cite{Andreou:2019ikc} suffices to show that the onset of scalarization is determined entirely by the coupling with ${\cal G}$. Interestingly, it has also been showed that black hole scalarization can be triggered by rapid rotation as well \cite{Dima:2020yac} and some corresponding scalarized black hole solutions have been found in Refs.~\cite{Herdeiro:2020wei,Berti:2020kgk}. In this scenario the threshold  is still controlled by ${\cal G}$; however, the critical contribution does not come from the mass of the black hole, but from the spin.
 
 As has been stressed in Ref.~\cite{Silva:2018qhn}, although the onset of scalarization is determined by terms that are linear (in the equations) in the scalar, the properties of the scalarized object depend crucially on nonlinear interactions, as these are the ones that quench the linear instability and determine its endpoint. Non-linearities can originate from scalar self-interactions \cite{Macedo:2019sem}, from the coupling function to ${\cal G}$ \cite{Doneva:2017bvd}, and from the backreaction of the scalar onto the metric.
The potential coupling between the scalar field and the Ricci scalar, $R$, has mostly been disregarded in the case of black holes.

As mentioned earlier, this is entirely justified when studying the onset of scalarization, as GR black holes have a vanishing $R$. However, it is bound to have an effect on the properties of scalarized objects, as it will contribute to the nonlinear quenching of the tachyonic instability that leads to scalarization. Indeed, as soon as the scalar becomes nontrivial, $R$ will cease to be zero and it will contribute directly to the effective mass of the scalar. From an effective field theory (EFT) perspective there seems to be no justification to exclude such a coupling. Moreover, it has been shown in Ref.~\cite{Antoniou:2020nax} that this coupling makes GR a cosmological attractor and hence reconciles Gauss-Bonnet scalarization  with late-time cosmological observations. It has also been pointed out in Ref.~\cite{Ventagli:2020rnx} that this coupling can help suppress scalarization of neutron stars and hence evade the relevant constaints. 

Motivated by the above, we examine in this paper the role a coupling with the Ricci scalar can have on scalarized black holes. We consider the simplest model that contains a coupling with both the Ricci scalar and Gauss-Bonnet invariant, and we  study static, spherically symmetric black holes. We explore the region of existence of scalarized solutions when varying both couplings and the black hole mass. We examine the influence of the Ricci coupling  on the scalar charge of the black holes, which is the quantity that controls the deviations from GR in the observation of binaries. We also discuss the role this coupling can play in stability considerations and in rendering black hole scalarization compatible with cosmological observations and strong gravity constraints from neutron stars. 

The structure of the paper is as follows: in Sec.~\ref{sec:setup} we introduce the model that we study. We discuss the scalarization thresholds and present the equations that one needs to solve to obtain solutions that describe static, spherically symmetric black holes. We then solve the equations perturbatively near the horizon and asymptotically far, and present the numerical implementation for solving the equations non-perturbatively. In Sec.~\ref{sec:numerical}, we present the solutions that we obtain numerically and fully explore their properties. Section~\ref{sec:discussion} contains further discussion on our results and future prospects.

\section{Setup}
\label{sec:setup}

We will consider the following action

\begin{equation}
\label{eq:ActionGeneric}
\begin{split}
    S=\frac{1}{2\kappa}\int\dd^4x\sqrt{-g}\bigg[& R-\dfrac12(\partial\phi)^2
    %\\&
    -\left(\frac{\beta}{2}R-\alpha\gb\right)\frac{\phi^2}{2} \bigg],%+S_\mathrm{M}.
\end{split}
\end{equation}
where $\kappa = 8\pi G/c^4$, $\beta$ is a dimensionless parameter, while $\alpha$ has dimensions of length squared.  The normalization of $\beta$ is chosen to match the standard DEF literature.

One can consider the above action as part of an EFT in which the scalar enjoys $\phi \to -\phi$ symmetry while shift symmetry is broken only by the coupling to gravity. For linear perturbations around solutions that solve the vacuum Einstein equations, the $\phi^2 {\cal G}$ coupling will be the leading correction to GR \cite{Andreou:2019ikc}. However, more generally, the $\phi^2 R$ term comes with a lower mass dimension and provides a direct contribution to the effective mass. Hence it is expected to play a crucial role in the non-linear quenching of the tachyonic instability that one associates with scalarization, and in determining the properties of scalarized black holes. The complete EFT would include more terms that can contribute to the effective mass (nonlinearly), such as the operators $R\phi^4$ and $G^{\mu\nu}\partial_\mu\phi\partial_\nu\phi$. These operators would enlarge the parameter space, while they are characterized by a higher mass dimension than $\phi^2 R$. We will neglect them in our analysis and we do not expect them to change the final results qualitatively. The modified Einstein equation is
%%%%
\begin{equation}\label{eq:grav_eq}
    G_{\mu\nu}=T^\phi_{\mu\nu},%+T^\text{M}_{\mu\nu},
\end{equation}
%%%%
where 
\begin{equation}
\label{eq:EMtensor}
\begin{split}
    T^\phi_{\mu\nu}=&-\frac{1}{4}g_{\mu\nu}(\nabla\phi)^2+\frac{1}{2}\nabla_\mu\phi\nabla_\nu\phi\\
    &+\frac{\beta\phi^2}{4}G_{\mu\nu}+\frac{\beta}{4}\left(g_{\mu\nu} \nabla^2 -\nabla_\mu\nabla_\nu \right)\phi^2\\
    &-\frac{\alpha}{4 g}g_{\mu(\rho}g_{\sigma)\nu}\epsilon^{\kappa\rho\alpha\beta}\epsilon^{\sigma\gamma\lambda\tau}R_{\lambda\tau\alpha\beta}\nabla_{\gamma}\nabla_{\kappa}\phi^2
\end{split}
\end{equation}
is the energy momentum tensor contribution that comes from the variation of the $\phi$-dependent part of the action with respect to the metric. The scalar field equation reads
%%%%
\begin{equation}\label{eq:scal_eq}
    \Box \phi =m_\text{eff}^2\phi,
\end{equation}
%%%%
where the effective scalar mass is given by
%%%%
\begin{equation}
    \meff^2=\frac{\beta}{2}R-\alpha \gb.
\end{equation}

\subsection{Scalarization threshold}\label{subsec:threshold}

\begin{figure}[t]
    \centering
    \includegraphics[width=0.45\textwidth]{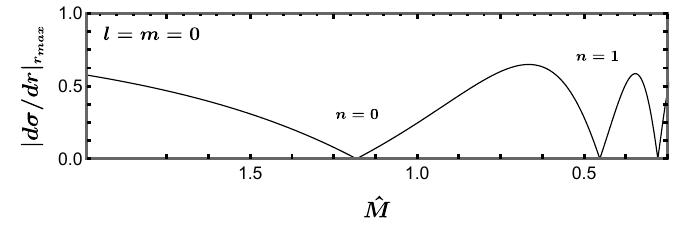}
    \caption{Numerical solution of the decoupled scalar equation on a GR background. The points where the line touches the horizontal axis correspond to the scalarization thresholds $\hat{M}^{(n)}_{\text{th}}$: for $\hat M<\hat{M}^{(n)}_{\text{th}}$, GR black holes are unstable under scalar perturbations with $n$ nodes. Note that the horizontal axis is inverted.}
    \label{fig:fig_Mth}
\end{figure}

As mentioned above, in linearized theory, scalarization manifests itself as a tachyonic instability around a GR solution.  Linearizing around a Schwarzschild background and neglecting backreaction,  we can recast the scalar equation \eqref{eq:scal_eq} into the following form:

\begin{equation}\label{eq:sigma}
    -\frac{\partial^2 \sigma}{\partial t^2}+\frac{\partial^2 \sigma}{\partial r_*^2}= V_{\text{eff}}\,\sigma,
\end{equation}
%%%%
where the scalar field is decomposed into spherical harmonics, $\delta \phi = \epsilon\, \sigma(r,t) Y^m_l(\theta,\varphi)/r$, $\dd r=\dd r_* e^{(\Gamma-\Lambda)/2}$, and 
\begin{equation}
    V_{\text{eff}}=e^\Gamma \left[\frac{e^{-\Lambda}}{2r}(\Gamma'-\Lambda')-\alpha\, \gb \right].
\end{equation}
The coupling between the scalar and $R$ does not contribute at all, as $R=0$ on a Schwarzschild background.
The effective potential (and hence the onset of scalarization) is controlled only by the rescaled mass $\hat M =M/\sqrt{\alpha}$, where $M$ is the mass of the black hole. GR solutions will become unstable for small values of $\hat M$, which correspond to larger curvatures or larger  $\alpha$ couplings. Unstable modes for $\phi$ will correspond to bound states for  $V_\text{eff}$ with $\delta\phi_{\infty}=0$. We search for such bound states while varying $\hat M$ and identify the threshold rescaled masses,  $\hat{M}^{(n)}_{\text{th}}$, ($n=0,1,2,$ etc). This is shown in Fig.~\ref{fig:fig_Mth}.
%%%%
 The mode associated with a threshold mass $\hat{M}^{(n)}_{\text{th}}$ has $n$ nodes. Hence, whenever $\hat M<\hat{M}^{(n)}_{\text{th}}$, GR black holes become unstable to a perturbation with $n$ nodes. Numerically, these thresholds are $(n,\hat{M}^{(n)}_{\text{th}})\approx (0,1.18),\, (1,0.45),\, (2,0.28)$, etc. They are in agreement with the results of \cite{Silva:2017uqg} under an appropriate rescaling.

\subsection{Static, spherically symmetric black holes}

We consider a static and spherically symmetric background:
\begin{equation}{\dd s}^2=-e^{\Gamma(r)}{\dd t}^2+e^{\Lambda(r)} {\dd r}^2
+r^2{\dd \Omega}^2.
\end{equation}
The field equations can be
cast as three coupled ordinary differential equations. The $(rr)$ component of the metric equations can be solved algebraically with respect to $e^\Lambda$:
%%%%
\begin{equation}\label{eq:grr}
    e^\Lambda= \frac{-B+\delta\sqrt{B^2-4A\,C}}{2A},\,\delta=\pm 1,
\end{equation}
%%%%
where
\begin{align}
    A=&\;4-\beta  \phi^2,\\
    \begin{split}
    B=&\; \beta  \phi ^2+\Gamma ' (\beta  r^2 \phi  \phi '-8 \alpha  \phi\,  \phi '+\beta  r \phi ^2-4 r)\\
    &+r^2 \phi '^2+4 \beta  r \phi\,  \phi '-4,
    \end{split}
    \\
    C=&\; 24 \alpha  \Gamma ' \phi\, \phi '.
\end{align}
%%%%
and a prime denotes differentiation with respect to the radial coordinate. By substituting \eqref{eq:grr} in the remaining field equations, we end up with a system of two coupled second order differential equations:
%%%%
\begin{align}
    \Gamma''=&\, \tilde{\Gamma}(r,\Gamma',\phi,\phi',\alpha,\beta)\label{eq:dif_1},\\
    \phi''=&\, \tilde{\phi}(r,\Gamma',\phi,\phi',\alpha,\beta)\label{eq:dif_2}.
\end{align}
%%%%

In order to search for black hole solutions, we assume the existence of a horizon, where $e^\Gamma\rightarrow 0,\; e^\Lambda\rightarrow \infty$. In line with previous results for different models fashioning a coupling with ${\cal G}$ ({\em e.g.}~\cite{Kanti:1995vq,Sotiriou:2014pfa}), only $\delta=+1$ leads to black hole solutions.

\subsection{Near-horizon expansion}

Near the horizon, one can perform the following expansion:
%%%%
\begin{align}
    e^\Gamma(r\approx r_\text{h})= & \,\gamma_1 (r-r_\text{h})+\gamma_2 (r-r_\text{h})^2+...\label{eq:exp_h1}\\
    e^{-\Lambda}(r\approx r_\text{h})= & \,\lambda_1 (r-r_\text{h})+\lambda_2 (r-r_\text{h})^2+...\label{eq:exp_h2}\\
    \phi(r\approx r_\text{h})= & \,\phi_\text{h} +\phi_1 (r-r_\text{h}) +\phi_2 (r-r_\text{h})^2+...\label{eq:exp_h3}
\end{align}
%%%%
One can substitute these expressions in Eqs.~\eqref{eq:grr}, \eqref{eq:dif_1} and \eqref{eq:dif_2}, and obtain a near-horizon solution. In particular, $\phi''_\text{h}$ remains finite only provided that
%%%%
\begin{equation}\label{eq:phi_h}
         \phi'(r_\text{h})=\phi_1= \big(a+\sqrt{\Delta}\big)/b,
\end{equation}
%%%%
where the expressions for $a,\;b$ and $\Delta$ are as follows:
\begin{align}
    a &=24 \alpha  \beta  r_h \phi _h^2+r_h^3 \left(-3 \beta ^2 \phi _h^2+\beta  \phi _h^2-4\right),
    \\[4mm]
    \begin{split}
    \Delta &=9216 \alpha ^3 \beta  \phi _h^4+r_h^6 \left(3 \beta ^2 \phi _h^2-\beta  \phi _h^2+4\right)^2\\
    &\quad -192 \alpha ^2 r_h^2 \phi _h^2 \left(9 \beta ^2 \phi _h^2-2 \beta  \phi _h^2+8\right),
    \end{split}
    \\[4mm]
    \begin{split}
    b &= 2 \phi _h \left(8 \alpha -\beta  r_h^2\right) \big[24 \alpha  \beta  \phi _h^2\\
    &\quad +r_h^2 \left(-3 \beta ^2 \phi _h^2+\beta  \phi _h^2-4\right)\big]/(\beta \phi_h^2-4).
    \end{split}
\end{align}

Requiring that $\Delta\geq0$  defines a region on the $(r_\text{h},\phi_\text{h})$ space where regular black hole solutions with scalar hair can be found.

\subsection{Asymptotic expansion}

In order to analyze the asymptotic behaviour of the solutions, one can perform a suitable expansion, and solve the equations near spatial infinity imposing that $\phi$ vanishes there. This yields 
%%%%
\begin{widetext}

\begin{align}
    \begin{split}\label{eq:exp_f1}
        g_{tt}(r\gg r_\text{h})= &\; 1-2M\big/{r}+\beta \, Q^2\big/{4\,r^2} +\big(M Q^2-3 \beta  M Q^2\big)\big/{12\, r^3}\\
        &+\big(8 M^2 Q^2-28 \beta  M^2 Q^2-3 \beta ^3 Q^4+5 \beta ^2 Q^4-\beta  Q^4\big)\big/{48\, r^4}+\big(288 M^3 Q^2-1040 \beta  M^3 Q^2\\
        &+3072 \alpha  M Q^2-60 \beta ^3 M Q^4+115 \beta ^2 M Q^4+10 \beta  M Q^4-9 M Q^4\big)\big/{960r^5}+\mathcal{O}\left(1/r^6\right) \;,
    \end{split}\\[4mm]
    \begin{split}\label{eq:exp_f2}
        g_{rr}(r\gg r_\text{h})= &\; 1+2M\big/r+\big(16 M^2+2 \beta \, Q^2-Q^2\big)\big/{4\,r^2}+\big(32 M^3-5 M Q^2+11 \beta  M Q^2\big)\big/{4\, r^3}\\
        &+\big(488 \beta  M^2 Q^2-208 M^2 Q^2+768 M^4-12 \beta ^3 Q^4+17 \beta ^2 Q^4-13 \beta  Q^4+3 Q^4\big)\big/{48\, r^4}\\
        &+\big(6064 \beta  M^3 Q^2-2464 M^3 Q^2+6144 M^5-1536 \alpha  M Q^2-348 \beta ^3 M Q^4+589 \beta ^2 M Q^4\\
        &-442 \beta  M Q^4+97
        M Q^4\big)\big/{192\, r^5} +\mathcal{O}\left(1/r^6\right) \;,
    \end{split}\\[4mm]
    \begin{split}\label{eq:exp_f3}
        \phi(r\gg r_\text{h})= &\; Q\big/r+M Q\big/r^2+\big(32 M^2 Q-3 \beta ^2 Q^3+2 \beta  Q^3-Q^3\big)\big/24\, r^3\\
        &+\big(48 M^3 Q-9 \beta ^2 M Q^3+9 \beta  M Q^3-4 M Q^3)\big/24\,r^4+\big(2240 \beta  M^2 Q^3-1680 \beta ^2 M^2 Q^3\\
        &-928 M^2 Q^3-4608 \alpha  M^2 Q+6144 M^4 Q+117 \beta ^4 Q^5-144 \beta ^3
        Q^5\\
        &+86 \beta ^2 Q^5-40 \beta  Q^5+9 Q^5\big)\big/{1920 \, r^5}+\mathcal{O}\left(1/r^6\right) \;,
    \end{split}
\end{align}

\end{widetext}
%%%%
where $M$ is the ADM mass and $Q$ is the scalar charge (although $Q$ is not associated to a conservation law, it does determine the decay of the scalar field at large distance). Equations \eqref{eq:exp_f1}-\eqref{eq:exp_f3} suggest, as one would expect, that the Ricci coupling dominates over the Gauss-Bonnet coupling at large radii. Specifically, the Ricci coupling appears at order $r^{-2}$, whereas the Gauss-Bonnet coupling appears initially at order $r^{-5}$.

\subsection{Numerical implementation}

The system of ordinary differential equations (ODEs) \eqref{eq:dif_1} and \eqref{eq:dif_2} can, in principle, be solved by starting from the horizon and integrating towards larger radii.   $\alpha$ and $\beta$ are theoretical parameters that are considered fixed. 
The values of $\Gamma'$, $\Gamma$, $\phi'$, and $\phi$ at $r=r_\text{h}$ appear to be ``initial data''. However, they are not all free to choose. $\Gamma(r_\text{h})$ is fixed by the condition $e^{\Gamma(r_\text{h})}=0$, {\em i.e.}~the fact that $r=r_\text{h}$ is a horizon. $\Gamma'$ has to diverge at $r=r_\text{h}$, else $e^{\Gamma}$ will have a vanishing derivative on the horizon. Finally, $\phi'(r_\text{h})$, and $\phi(r_\text{h})$ are related by the regularity condition \eqref{eq:phi_h}.  
One also needs to fix $r_\text{h}$. The field equations are invariant under the global scaling symmetry $r\to\mu r$, $\alpha\to\mu^2\alpha$, where $\mu$ is a free parameter. We can make use of this symmetry to reduce the space of parameters that we have to explore. Practically, we can decide that the horizon is located at $r_\text{h}=1$; solutions with $r_\text{h}\neq1$ can later be obtained by a global scaling.

Hence, one can treat $\phi(r_\text{h})=\phi_\text{h}$ as the only free parameter. Integrating outwards, one will generically find a solution for arbitrary $\phi_\text{h}$. However, for given $\alpha$ and $\beta$, only one value of $\phi_\text{h}$ has the desired asymptotics, namely $\phi(r\to \infty)=\phi_\infty=0$. Imposing this condition (by a shooting method and to a desired precision) yields a unique solution. The global rescaling mentioned above turns this solution into a one-parameter family, that we can interpret as a family of black holes parametrized by their ADM mass $M$, for fixed couplings $\alpha$ and $\beta$. The scalar charge $Q$ is then determined as a function of $M$, $\alpha$ and $\beta$. 

A practical complication is that the regularity condition of Eq.~\eqref{eq:phi_h} cannot be imposed numerically with any reasonable accuracy. To circumvent this problem we start the numerical integration at $r\approx r_\text{h}[1 + \mathcal{O}(10^{-4})]$ and use the perturbative expansion in Eqs.~\eqref{eq:exp_h1}--\eqref{eq:exp_h3} to impose the regularity and propagate the data from the horizon to the starting point of the numerical integration. We typically integrate up to distances $r/r_\text{h}\approx 10^4$ and impose that $\phi$ vanishes there to a part in $10^{4}$.

In the next section, we will be using scale-invariant masses and charges, defined as
%%%%
\begin{equation}
\label{eq:scalinvMQ}
    \hat{M}=M/\sqrt{\alpha}\,,\;\; \hat{Q}=Q/\sqrt{\alpha}.
\end{equation}
Equation \eqref{eq:scalinvMQ} assumes that $\alpha>0$. 
Indeed, we will restrict our analysis to positive values of $\alpha$. Evading the no-hair theorem of Ref.~\cite{Silva:2017uqg} requires $\alpha>0$ when $\beta=0$ and $\gb$ is positive, which is the case for a Schwarzschild black hole. Moreover, the Ricci coupling, controlled by $\beta$, does not contribute to linear perturbation theory around GR black holes.  It is hence unlikely that scalarized spherically symmetric black hole solutions will exist for $\alpha<0$. It should be stressed, however, that the $\alpha<0$ case is  particularly interesting when studying rotating black holes \cite{Dima:2020yac}.

\section{Numerical Results}\label{sec:numerical}
\begin{figure*}[t]
\begin{center}
    \includegraphics[scale=0.72]{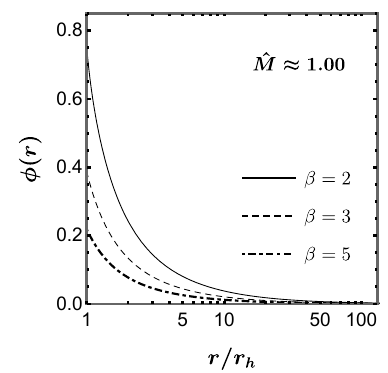}
    \includegraphics[scale=0.72]{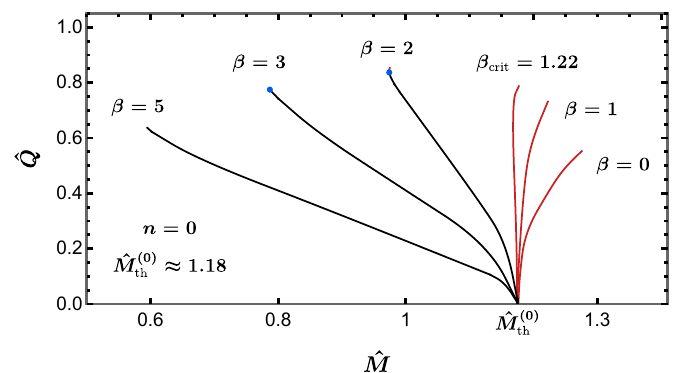}
    \includegraphics[scale=0.72]{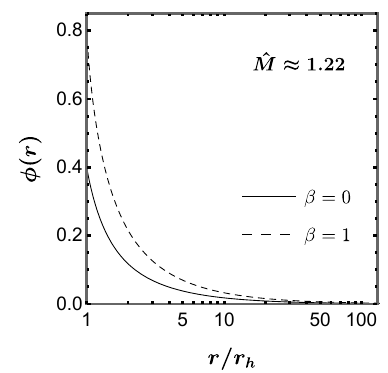}
\end{center}
\caption{(Left) Scalar field profile versus the normalized distance from the horizon for a sample mass $\hat{M}<\hat{M}^{(0)}_\text{th}$, and zero nodes for the scalar field radial profile. (Centre) Normalized scalar charge versus the normalized ADM mass for $n=0$ nodes solutions. The black lines correspond to values of $\beta$ for which all scalarized black holes have masses below the GR instability mass threshold, while the red lines mark values of $\beta$ that lead to scalarized black hole masses that are larger than the GR mass threshold. The blue dots mark the existence of a turning point, past which solutions are expected to be unstable. (Right) Scalar field profile versus the normalized distance from the horizon for a sample mass $\hat{M}>\hat{M}^{(0)}_\text{th}$.}
\label{fig:fig_2}
\end{figure*}

\subsection{$\boldsymbol{n=0}$ nodes for the scalar profile}

\begin{figure*}[ht]
\begin{center}
    \includegraphics[width=0.45\textwidth]{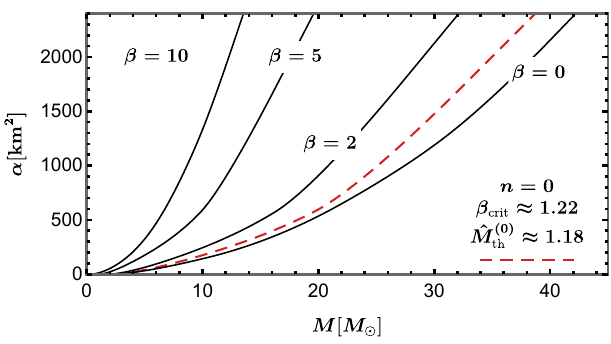}\hspace{1cm}
    \includegraphics[width=0.45\textwidth]{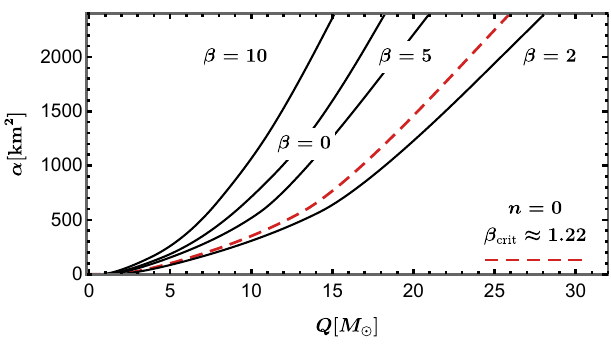}
\end{center}
\caption{(Left) Domain of existence of $n=0$ scalarized black holes on the $\alpha-M$ plane. For a given $\beta$, solutions exist between the corresponding black line, and the dashed, red line. The latter coincides with the line where GR solutions of equal mass would become unstable. (Right) Same but on the $\alpha-Q$ plane. Both panels can be obtained from an ``unfolding'' of Fig.~\ref{fig:fig_2}.}
\label{fig:fig_aMQdomain}
\end{figure*}

The first scenario we examine is the one where $\beta>0$. This scenario is motivated by the results of Ref.~\cite{Antoniou:2020nax}, where it was shown that positive values of $\beta$ make GR a cosmological attractor. We start by exploring the solutions characterized by $n=0$. The results are summarized in Fig.~\ref{fig:fig_2} and  Fig.~\ref{fig:fig_aMQdomain}.
The central plot of Fig.~\ref{fig:fig_2} shows the dependence of the scalar charge on the black hole ADM mass for different choices of $\beta$. When $\beta$  is smaller than some critical value $\beta_{\text{crit}}\approx 1.22$, the charge-mass curve tilts to the right and all scalarized black holes have larger ADM masses than the GR mass instability threshold. Such scalarized black holes are unlikely to be produced dynamically. ADM mass is a measure of energy for the system. The fact that {\em all} scalarized black holes for $\beta<\beta_{\text{crit}}$ have larger mass than {\em all} GR black holes that are unstable implies that, if {\em any} scalarized black hole is considered the end point of the tachyonic instability for a GR black hole, then this end state would have more energy than the initial state. 

Based on the argument above, we conjecture that scalarized black holes are unstable for $\beta<\beta_{\text{crit}}$. Conversely, for $\beta>\beta_{\text{crit}}$ the ADM mass for scalarized black holes can be smaller than the GR counterparts and hence it is reasonable to expect that scalarized black hole are endpoints of the tachyonic instability. These arguments are consistent with earlier results. In particular, it is already known that for $\beta=0$ scalarized black holes are radially unstable \cite{Blazquez-Salcedo:2018jnn}. Moreover, the general picture shown in the centre plot of Fig.~\ref{fig:fig_2} is very similar to the one presented in Ref.~\cite{Macedo:2019sem}. In that work, $\beta$ was vanishing and the $\phi^2 R$ term was absent, but a $\phi^4$ self-interaction had a similar effect. Analysis of radial stability did show in that case that stability was associated with whether the curves on the $\hat{Q}-\hat{M}$ plane tilt to the right or the left. 

These considerations suggest strongly that the coupling between $\phi$ and the Ricci scalar, can have  a very interesting stabilizing effect for scalarized black holes, without having to resort to scalar self-interactions. 

Note that in some cases, when $\beta>\beta_{\text{crit}}$ and hence the $\hat Q-\hat M$ curve initially leans to the left, this same curve later turns towards the right. The points at which the curves turn right are marked by blue dots in Fig.~\ref{fig:fig_2}. The right-leaning part of these curves is hardly noticeable in Fig.~\ref{fig:fig_2} because it is very short. One expects configurations past the turning point to be unstable, as configurations of the same ADM mass and smaller scalar charge exist. 

As is clear from Fig.~\ref{fig:fig_2}, for $\beta>\beta_{\text{crit}}$, the normalized scalar charge $\hat Q$ increases as the normalized ADM mass $\hat M$ decreases, at least in the part of the curves up to the turning point (blue dot), whereas for $\beta<\beta_{\text{crit}}$, the normalized scalar charge $\hat Q$ increases as the normalized ADM mass $\hat M$ increases. Interestingly, the dependence of the curvature near the horizon on the ADM mass turns out to be different in the two cases. For $\beta>\beta_{\text{crit}}$ scalarized black holes tend to have larger curvatures at the horizon when the ADM mass decreases, as is the case in GR, whereas for $\beta<\beta_{\text{crit}}$ the curvature on the horizon tends to increase as the mass (and scalar charge) increases. Hence, in both cases, the scalar charge seems to be controlled by the curvature.

In Fig.~\ref{fig:fig_aMQdomain}, we show the domain of existence of scalarized black holes on the $\alpha-M$ and $\alpha-Q$ planes. As discussed in Sec.~\ref{sec:setup}, linear analysis showed distinct scalarization thresholds, the first (zero nodes) of which we denote with $\hat{M}^{(0)}_{\text{th}}\approx 1.175$. This threshold  is represented by the dashed, red line in Fig.~\ref{fig:fig_aMQdomain}. Note that $\hat{M}=\text{constant}$ (respectively $\hat{Q}=\text{constant}$) translates to a parabola in the $\alpha-M$ ($\alpha-Q$) plane. The rest of the curves correspond to the existence boundaries for various values of $\beta$. They are related to the horizon condition presented in Eq.~\eqref{eq:phi_h}. Solutions then exist everywhere between the red, dashed GR instability line and the plain, black existence line. Examining the plots reveals something rather interesting: the value of the Ricci coupling $\beta$ can affect the relative position of the existence line with respect to the instability parabola. This should not come as a surprise, based on the results presented in Fig.~\ref{fig:fig_2}, where $\beta$ has a similar effect on the relative position of the curve with respect to the threshold mass $\hat{M}^{(0)}_{\text{th}}$.

As mentioned earlier, we do not plan to consider the $\beta<0$ case in any detail as positive values appear to be better motivated. However, we can report the following based on a preliminary exploration. There is still a critical value of $\beta$, and for $\beta$ smaller than this value, scalarized black holes have smaller ADM masses than the GR instability threshold, together with scalar charges that tend to increase with decreasing mass. For $\beta$ larger than the critical value, the behaviour is reversed. Hence, the equivalent to Fig.~\ref{fig:fig_2} would be qualitatively similar for $\beta<0$.

\subsection{$\boldsymbol{n=1,2}$ nodes for the scalar profile}

\begin{figure*}[t]
\begin{center}
    \includegraphics[width=0.45\textwidth]{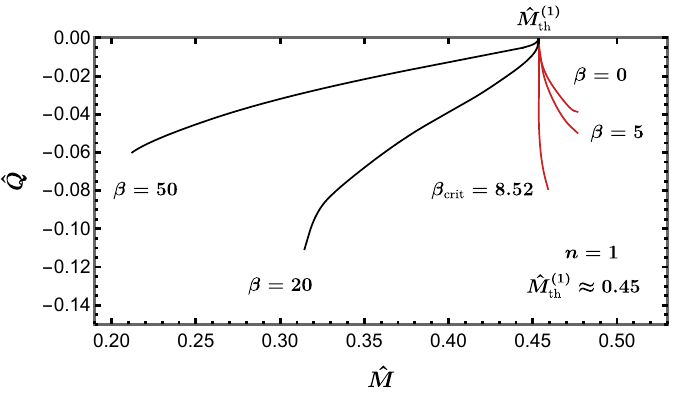}\hspace{1cm}
    \includegraphics[width=0.45\textwidth]{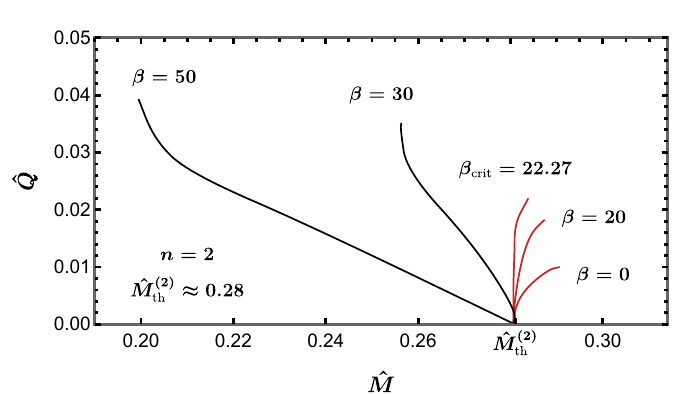}
\end{center}
\caption{Normalized scalar charge versus normalized ADM mass for the solutions with $n=1$ (left panel) and 2 (right panel) nodes.}
    \label{fig:fig_QM1}
\end{figure*}

We now turn to solutions characterized by $n=1$ and $n=2$. For  $\beta>0$, the plot of the normalized charge versus the normalized mass is given in Fig.~\ref{fig:fig_QM1}.
A noticeable pattern is that, for $n=0$, the scalar charge is positive, while it is negative for $n=1$, positive for $n=2$, and so on. This is simply due to the fact that the scalar field has to approach 0 at spatial infinity from a different side, depending on the number of nodes. There is no deep significance in the sign of the charge, since the action \eqref{eq:ActionGeneric} possesses parity symmetry $\phi\to-\phi$, and the signs would have been flipped, had we chosen negative values of $\phi_\text{h}$ as initial conditions. Compared to the $n=0$ case, we see that the order of magnitude of the charge for all different values of $\beta$ is significantly smaller, and the range of masses for which we find scalarized solutions is strongly reduced. Once again there is a critical value of $\beta$ that separates right-leaning curves (likely unstable) from left leaning ones (likely stable).

\section{Discussion}
\label{sec:discussion}

We have considered the contribution of a coupling between the scalar, $\phi$, and the Ricci scalar, $R$, can have in black hole scalarization. We focused on static, spherically symmetric black holes. The $\beta \phi^2 R$ coupling is known not to affect the threshold of scalarization. However, our results show that it can  alter the domain of existence of scalarized black holes, significantly modify their properties, and control their scalar charge. Our results also strongly suggest that the strength of this coupling can have an impact on the stability of scalarized black holes. In particular, having $\beta$ be larger than some critical value, $\beta_\text{crit}$, is expected to resolve the stability problems for models that do not include the $\beta \phi^2 R$ coupling. We will investigate this issue in more detail in future work. 

We have mostly focused on positive values for $\beta$.  We did so for two reasons. First, it has recently been shown that including the $\beta \phi^2 R$ term in black hole scalarization models and selecting a positive $\beta$ makes GR a cosmological attractor and allows one to have a consistent cosmological history, at least from the end of the inflationary era \cite{Antoniou:2020nax}.  The numerical values that we considered here for the couplings are similar to those used in \cite{Antoniou:2020nax}.  Second, for positive values of the Ricci coupling (and reasonably small values of the Gauss-Bonnet coupling), neutron stars do not scalarize \cite{Ventagli:2020rnx}. This allows one to evade the very tight binary pulsar constraints ({\em e.g.} \cite{Freire:2012mg,Antoniadis:2013pzd,Shao:2017gwu}), related to energy losses due to dipolar emission of gravitational waves, without the need to add a bare mass to the scalar (and tune it appropriately). 

It is clear that inclusion of the $\beta \phi^2 R$ coupling has multiple benefits in scalarization models. It is worth re-iterating that this coupling has lower mass dimensions than the $\alpha \phi^2 {\cal G}$ coupling, which triggers scalarization at linear level. Moreover, unlike a bare mass term or scalar self-interactions, it allows the scalar to remain massless and free in flat space. Hence, the $\beta \phi^2 R$ coupling can be part of an interesting EFT that respects $\phi \to -\phi$ symmetry and in which shift symmetry can be broken only via the coupling to gravity (the complete EFT would potentially include more terms, such as  $R\phi^4$ and $G^{\mu\nu}\partial_\mu\phi\partial_\nu\phi$).

Gravitational wave observation of binaries that contain black holes would still be able to measure or constrain $\beta$ and $\alpha$. A detailed post-Newtonian analysis of the inspiral phase would be sufficient to provide some first constraints. Scalarization models in which the scalar charge is non-zero only below a mass threshold are also expected to be severely constrained by extreme mass ratio inspirals (EMRIs) observations by LISA: the supermassive black hole would be described by the Kerr metric, whereas the small black hole can carry a scalar charge. This is the ideal scenario to apply the considerations of Ref.~\cite{Maselli:2020zgv}.

It should be stressed that we only considered the case $\alpha>0$ throughout this paper, as this is a requirement for having scalarized black holes under the assumptions of staticity and spherical symmetry. However, it has been shown in Ref.~\cite{Dima:2020yac} that, for $\alpha<0$ (and $\beta=0$), scalarization can be triggered by rapid rotation. Indeed, some scalarized black hole have been found in this scenario in Refs.~\cite{Herdeiro:2020wei,Berti:2020kgk}. It would thus be very interesting to consider the effect of the $\beta \phi^2 R$ coupling for $\alpha<0$, {\em i.e.}~in models where scalarization is induced by rotation. 

It is likely that theoretical constraints on the value of couplings in scalarization models will be imposed by the requirement that the initial value problem be well-posed in dynamical evolution scenarios where one expects the models to be good EFTs. Results in this direction have been obtained in Ref.~\cite{Ripley:2020vpk} for $\beta=0$. The inclusion of the coupling with the Ricci scalar is likely to affect the results quantitatively, and hence is an interesting prospect.

\begin{acknowledgments}
 G.A. acknowledges partial support from
the Onassis Foundation.
This project has received funding from the European Union's Horizon 2020 research and innovation programme under the Marie Sklodowska-Curie grant agreement No 101007855.
A.L. thanks FCT for financial support through Project~No.~UIDB/00099/2020.
A.L. acknowledges financial support provided by FCT/Portugal through grants PTDC/MAT-APL/30043/2017 and PTDC/FIS-AST/7002/2020.
T.P.S. acknowledges partial support from the STFC Consolidated Grant No. ST/T000732/1. 
We also acknowledge  networking support by the GWverse COST Action
CA16104, ``Black holes, gravitational waves and fundamental physics.''
\end{acknowledgments}

\bibliography{bibnote}

\begin{thebibliography}{34}%
\makeatletter
\providecommand \@ifxundefined [1]{%
 \@ifx{#1\undefined}
}%
\providecommand \@ifnum [1]{%
 \ifnum #1\expandafter \@firstoftwo
 \else \expandafter \@secondoftwo
 \fi
}%
\providecommand \@ifx [1]{%
 \ifx #1\expandafter \@firstoftwo
 \else \expandafter \@secondoftwo
 \fi
}%
\providecommand \natexlab [1]{#1}%
\providecommand \enquote  [1]{``#1''}%
\providecommand \bibnamefont  [1]{#1}%
\providecommand \bibfnamefont [1]{#1}%
\providecommand \citenamefont [1]{#1}%
\providecommand \href@noop [0]{\@secondoftwo}%
\providecommand \href [0]{\begingroup \@sanitize@url \@href}%
\providecommand \@href[1]{\@@startlink{#1}\@@href}%
\providecommand \@@href[1]{\endgroup#1\@@endlink}%
\providecommand \@sanitize@url [0]{\catcode `\\12\catcode `\$12\catcode
  `\&12\catcode `\#12\catcode `\^12\catcode `\_12\catcode `\%12\relax}%
\providecommand \@@startlink[1]{}%
\providecommand \@@endlink[0]{}%
\providecommand \url  [0]{\begingroup\@sanitize@url \@url }%
\providecommand \@url [1]{\endgroup\@href {#1}{\urlprefix }}%
\providecommand \urlprefix  [0]{URL }%
\providecommand \Eprint [0]{\href }%
\providecommand \doibase [0]{http://dx.doi.org/}%
\providecommand \selectlanguage [0]{\@gobble}%
\providecommand \bibinfo  [0]{\@secondoftwo}%
\providecommand \bibfield  [0]{\@secondoftwo}%
\providecommand \translation [1]{[#1]}%
\providecommand \BibitemOpen [0]{}%
\providecommand \bibitemStop [0]{}%
\providecommand \bibitemNoStop [0]{.\EOS\space}%
\providecommand \EOS [0]{\spacefactor3000\relax}%
\providecommand \BibitemShut  [1]{\csname bibitem#1\endcsname}%
\let\auto@bib@innerbib\@empty
%</preamble>
\bibitem [{\citenamefont {Abbott}\ \emph {et~al.}(2016)\citenamefont {Abbott}
  \emph {et~al.}}]{Abbott:2016blz}%
  \BibitemOpen
  \bibfield  {author} {\bibinfo {author} {\bibfnamefont {B.~P.}\ \bibnamefont
  {Abbott}} \emph {et~al.} (\bibinfo {collaboration} {LIGO Scientific,
  Virgo}),\ }\href {\doibase 10.1103/PhysRevLett.116.061102} {\bibfield
  {journal} {\bibinfo  {journal} {Phys. Rev. Lett.}\ }\textbf {\bibinfo
  {volume} {116}},\ \bibinfo {pages} {061102} (\bibinfo {year} {2016})},\
  \Eprint {http://arxiv.org/abs/1602.03837} {arXiv:1602.03837 [gr-qc]}
  \BibitemShut {NoStop}%
\bibitem [{\citenamefont {Abbott}\ \emph {et~al.}(2017)\citenamefont {Abbott}
  \emph {et~al.}}]{TheLIGOScientific:2017qsa}%
  \BibitemOpen
  \bibfield  {author} {\bibinfo {author} {\bibfnamefont {B.~P.}\ \bibnamefont
  {Abbott}} \emph {et~al.} (\bibinfo {collaboration} {LIGO Scientific,
  Virgo}),\ }\href {\doibase 10.1103/PhysRevLett.119.161101} {\bibfield
  {journal} {\bibinfo  {journal} {Phys. Rev. Lett.}\ }\textbf {\bibinfo
  {volume} {119}},\ \bibinfo {pages} {161101} (\bibinfo {year} {2017})},\
  \Eprint {http://arxiv.org/abs/1710.05832} {arXiv:1710.05832 [gr-qc]}
  \BibitemShut {NoStop}%
%%CITATION = ARXIV:1710.05832;%%
\bibitem [{\citenamefont {Abbott}\ \emph {et~al.}(2020)\citenamefont {Abbott}
  \emph {et~al.}}]{Abbott:2020niy}%
  \BibitemOpen
  \bibfield  {author} {\bibinfo {author} {\bibfnamefont {R.}~\bibnamefont
  {Abbott}} \emph {et~al.} (\bibinfo {collaboration} {LIGO Scientific,
  Virgo}),\ }\href@noop {} {\  (\bibinfo {year} {2020})},\ \Eprint
  {http://arxiv.org/abs/2010.14527} {arXiv:2010.14527 [gr-qc]} \BibitemShut
  {NoStop}%
\bibitem [{\citenamefont {Barack}\ \emph {et~al.}(2019)\citenamefont {Barack}
  \emph {et~al.}}]{Barack:2018yly}%
  \BibitemOpen
  \bibfield  {author} {\bibinfo {author} {\bibfnamefont {L.}~\bibnamefont
  {Barack}} \emph {et~al.},\ }\href {\doibase 10.1088/1361-6382/ab0587}
  {\bibfield  {journal} {\bibinfo  {journal} {Class. Quant. Grav.}\ }\textbf
  {\bibinfo {volume} {36}},\ \bibinfo {pages} {143001} (\bibinfo {year}
  {2019})},\ \Eprint {http://arxiv.org/abs/1806.05195} {arXiv:1806.05195
  [gr-qc]} \BibitemShut {NoStop}%
\bibitem [{\citenamefont {Sathyaprakash}\ \emph {et~al.}(2019)\citenamefont
  {Sathyaprakash} \emph {et~al.}}]{Sathyaprakash:2019yqt}%
  \BibitemOpen
  \bibfield  {author} {\bibinfo {author} {\bibfnamefont {B.~S.}\ \bibnamefont
  {Sathyaprakash}} \emph {et~al.},\ }\href@noop {} {\  (\bibinfo {year}
  {2019})},\ \Eprint {http://arxiv.org/abs/1903.09221} {arXiv:1903.09221
  [astro-ph.HE]} \BibitemShut {NoStop}%
\bibitem [{\citenamefont {Barausse}\ \emph {et~al.}(2020)\citenamefont
  {Barausse} \emph {et~al.}}]{Barausse:2020rsu}%
  \BibitemOpen
  \bibfield  {author} {\bibinfo {author} {\bibfnamefont {E.}~\bibnamefont
  {Barausse}} \emph {et~al.},\ }\href {\doibase 10.1007/s10714-020-02691-1}
  {\bibfield  {journal} {\bibinfo  {journal} {Gen. Rel. Grav.}\ }\textbf
  {\bibinfo {volume} {52}},\ \bibinfo {pages} {81} (\bibinfo {year} {2020})},\
  \Eprint {http://arxiv.org/abs/2001.09793} {arXiv:2001.09793 [gr-qc]}
  \BibitemShut {NoStop}%
\bibitem [{\citenamefont {Damour}\ and\ \citenamefont
  {Nordtvedt}(1993)}]{Damour:1992kf}%
  \BibitemOpen
  \bibfield  {author} {\bibinfo {author} {\bibfnamefont {T.}~\bibnamefont
  {Damour}}\ and\ \bibinfo {author} {\bibfnamefont {K.}~\bibnamefont
  {Nordtvedt}},\ }\href {\doibase 10.1103/PhysRevLett.70.2217} {\bibfield
  {journal} {\bibinfo  {journal} {Phys. Rev. Lett.}\ }\textbf {\bibinfo
  {volume} {70}},\ \bibinfo {pages} {2217} (\bibinfo {year}
  {1993})}\BibitemShut {NoStop}%
%%CITATION = PRLTA,70,2217;%%
\bibitem [{\citenamefont {Freire}\ \emph {et~al.}(2012)\citenamefont {Freire},
  \citenamefont {Wex}, \citenamefont {Esposito-Farese}, \citenamefont
  {Verbiest}, \citenamefont {Bailes}, \citenamefont {Jacoby}, \citenamefont
  {Kramer}, \citenamefont {Stairs}, \citenamefont {Antoniadis},\ and\
  \citenamefont {Janssen}}]{Freire:2012mg}%
  \BibitemOpen
  \bibfield  {author} {\bibinfo {author} {\bibfnamefont {P.~C.~C.}\
  \bibnamefont {Freire}}, \bibinfo {author} {\bibfnamefont {N.}~\bibnamefont
  {Wex}}, \bibinfo {author} {\bibfnamefont {G.}~\bibnamefont
  {Esposito-Farese}}, \bibinfo {author} {\bibfnamefont {J.~P.~W.}\ \bibnamefont
  {Verbiest}}, \bibinfo {author} {\bibfnamefont {M.}~\bibnamefont {Bailes}},
  \bibinfo {author} {\bibfnamefont {B.~A.}\ \bibnamefont {Jacoby}}, \bibinfo
  {author} {\bibfnamefont {M.}~\bibnamefont {Kramer}}, \bibinfo {author}
  {\bibfnamefont {I.~H.}\ \bibnamefont {Stairs}}, \bibinfo {author}
  {\bibfnamefont {J.}~\bibnamefont {Antoniadis}}, \ and\ \bibinfo {author}
  {\bibfnamefont {G.~H.}\ \bibnamefont {Janssen}},\ }\href {\doibase
  10.1111/j.1365-2966.2012.21253.x} {\bibfield  {journal} {\bibinfo  {journal}
  {Mon. Not. Roy. Astron. Soc.}\ }\textbf {\bibinfo {volume} {423}},\ \bibinfo
  {pages} {3328} (\bibinfo {year} {2012})},\ \Eprint
  {http://arxiv.org/abs/1205.1450} {arXiv:1205.1450 [astro-ph.GA]} \BibitemShut
  {NoStop}%
\bibitem [{\citenamefont {Antoniadis}\ \emph {et~al.}(2013)\citenamefont
  {Antoniadis} \emph {et~al.}}]{Antoniadis:2013pzd}%
  \BibitemOpen
  \bibfield  {author} {\bibinfo {author} {\bibfnamefont {J.}~\bibnamefont
  {Antoniadis}} \emph {et~al.},\ }\href {\doibase 10.1126/science.1233232}
  {\bibfield  {journal} {\bibinfo  {journal} {Science}\ }\textbf {\bibinfo
  {volume} {340}},\ \bibinfo {pages} {6131} (\bibinfo {year} {2013})},\ \Eprint
  {http://arxiv.org/abs/1304.6875} {arXiv:1304.6875 [astro-ph.HE]} \BibitemShut
  {NoStop}%
\bibitem [{\citenamefont {Shao}\ \emph {et~al.}(2017)\citenamefont {Shao},
  \citenamefont {Sennett}, \citenamefont {Buonanno}, \citenamefont {Kramer},\
  and\ \citenamefont {Wex}}]{Shao:2017gwu}%
  \BibitemOpen
  \bibfield  {author} {\bibinfo {author} {\bibfnamefont {L.}~\bibnamefont
  {Shao}}, \bibinfo {author} {\bibfnamefont {N.}~\bibnamefont {Sennett}},
  \bibinfo {author} {\bibfnamefont {A.}~\bibnamefont {Buonanno}}, \bibinfo
  {author} {\bibfnamefont {M.}~\bibnamefont {Kramer}}, \ and\ \bibinfo {author}
  {\bibfnamefont {N.}~\bibnamefont {Wex}},\ }\href {\doibase
  10.1103/PhysRevX.7.041025} {\bibfield  {journal} {\bibinfo  {journal} {Phys.
  Rev.}\ }\textbf {\bibinfo {volume} {X7}},\ \bibinfo {pages} {041025}
  (\bibinfo {year} {2017})},\ \Eprint {http://arxiv.org/abs/1704.07561}
  {arXiv:1704.07561 [gr-qc]} \BibitemShut {NoStop}%
%%CITATION = ARXIV:1704.07561;%%
\bibitem [{\citenamefont {Ramazanoğlu}\ and\ \citenamefont
  {Pretorius}(2016)}]{Ramazanoglu:2016kul}%
  \BibitemOpen
  \bibfield  {author} {\bibinfo {author} {\bibfnamefont {F.~M.}\ \bibnamefont
  {Ramazanoğlu}}\ and\ \bibinfo {author} {\bibfnamefont {F.}~\bibnamefont
  {Pretorius}},\ }\href {\doibase 10.1103/PhysRevD.93.064005} {\bibfield
  {journal} {\bibinfo  {journal} {Phys. Rev.}\ }\textbf {\bibinfo {volume}
  {D93}},\ \bibinfo {pages} {064005} (\bibinfo {year} {2016})},\ \Eprint
  {http://arxiv.org/abs/1601.07475} {arXiv:1601.07475 [gr-qc]} \BibitemShut
  {NoStop}%
%%CITATION = ARXIV:1601.07475;%%
\bibitem [{\citenamefont {Cardoso}\ \emph
  {et~al.}(2013{\natexlab{a}})\citenamefont {Cardoso}, \citenamefont {Carucci},
  \citenamefont {Pani},\ and\ \citenamefont {Sotiriou}}]{Cardoso:2013opa}%
  \BibitemOpen
  \bibfield  {author} {\bibinfo {author} {\bibfnamefont {V.}~\bibnamefont
  {Cardoso}}, \bibinfo {author} {\bibfnamefont {I.~P.}\ \bibnamefont
  {Carucci}}, \bibinfo {author} {\bibfnamefont {P.}~\bibnamefont {Pani}}, \
  and\ \bibinfo {author} {\bibfnamefont {T.~P.}\ \bibnamefont {Sotiriou}},\
  }\href {\doibase 10.1103/PhysRevD.88.044056} {\bibfield  {journal} {\bibinfo
  {journal} {Phys. Rev.}\ }\textbf {\bibinfo {volume} {D88}},\ \bibinfo {pages}
  {044056} (\bibinfo {year} {2013}{\natexlab{a}})},\ \Eprint
  {http://arxiv.org/abs/1305.6936} {arXiv:1305.6936 [gr-qc]} \BibitemShut
  {NoStop}%
%%CITATION = ARXIV:1305.6936;%%
\bibitem [{\citenamefont {Cardoso}\ \emph
  {et~al.}(2013{\natexlab{b}})\citenamefont {Cardoso}, \citenamefont {Carucci},
  \citenamefont {Pani},\ and\ \citenamefont {Sotiriou}}]{Cardoso:2013fwa}%
  \BibitemOpen
  \bibfield  {author} {\bibinfo {author} {\bibfnamefont {V.}~\bibnamefont
  {Cardoso}}, \bibinfo {author} {\bibfnamefont {I.~P.}\ \bibnamefont
  {Carucci}}, \bibinfo {author} {\bibfnamefont {P.}~\bibnamefont {Pani}}, \
  and\ \bibinfo {author} {\bibfnamefont {T.~P.}\ \bibnamefont {Sotiriou}},\
  }\href {\doibase 10.1103/PhysRevLett.111.111101} {\bibfield  {journal}
  {\bibinfo  {journal} {Phys. Rev. Lett.}\ }\textbf {\bibinfo {volume} {111}},\
  \bibinfo {pages} {111101} (\bibinfo {year} {2013}{\natexlab{b}})},\ \Eprint
  {http://arxiv.org/abs/1308.6587} {arXiv:1308.6587 [gr-qc]} \BibitemShut
  {NoStop}%
%%CITATION = ARXIV:1308.6587;%%
\bibitem [{\citenamefont {Hawking}(1972)}]{Hawking:1972qk}%
  \BibitemOpen
  \bibfield  {author} {\bibinfo {author} {\bibfnamefont {S.~W.}\ \bibnamefont
  {Hawking}},\ }\href {\doibase 10.1007/BF01877518} {\bibfield  {journal}
  {\bibinfo  {journal} {Commun. Math. Phys.}\ }\textbf {\bibinfo {volume}
  {25}},\ \bibinfo {pages} {167} (\bibinfo {year} {1972})}\BibitemShut
  {NoStop}%
%%CITATION = CMPHA,25,167;%%
\bibitem [{\citenamefont {Sotiriou}\ and\ \citenamefont
  {Faraoni}(2012)}]{Sotiriou:2011dz}%
  \BibitemOpen
  \bibfield  {author} {\bibinfo {author} {\bibfnamefont {T.~P.}\ \bibnamefont
  {Sotiriou}}\ and\ \bibinfo {author} {\bibfnamefont {V.}~\bibnamefont
  {Faraoni}},\ }\href {\doibase 10.1103/PhysRevLett.108.081103} {\bibfield
  {journal} {\bibinfo  {journal} {Phys. Rev. Lett.}\ }\textbf {\bibinfo
  {volume} {108}},\ \bibinfo {pages} {081103} (\bibinfo {year} {2012})},\
  \Eprint {http://arxiv.org/abs/1109.6324} {arXiv:1109.6324 [gr-qc]}
  \BibitemShut {NoStop}%
%%CITATION = ARXIV:1109.6324;%%
\bibitem [{\citenamefont {Silva}\ \emph {et~al.}(2018)\citenamefont {Silva},
  \citenamefont {Sakstein}, \citenamefont {Gualtieri}, \citenamefont
  {Sotiriou},\ and\ \citenamefont {Berti}}]{Silva:2017uqg}%
  \BibitemOpen
  \bibfield  {author} {\bibinfo {author} {\bibfnamefont {H.~O.}\ \bibnamefont
  {Silva}}, \bibinfo {author} {\bibfnamefont {J.}~\bibnamefont {Sakstein}},
  \bibinfo {author} {\bibfnamefont {L.}~\bibnamefont {Gualtieri}}, \bibinfo
  {author} {\bibfnamefont {T.~P.}\ \bibnamefont {Sotiriou}}, \ and\ \bibinfo
  {author} {\bibfnamefont {E.}~\bibnamefont {Berti}},\ }\href {\doibase
  10.1103/PhysRevLett.120.131104} {\bibfield  {journal} {\bibinfo  {journal}
  {Phys. Rev. Lett.}\ }\textbf {\bibinfo {volume} {120}},\ \bibinfo {pages}
  {131104} (\bibinfo {year} {2018})},\ \Eprint
  {http://arxiv.org/abs/1711.02080} {arXiv:1711.02080 [gr-qc]} \BibitemShut
  {NoStop}%
%%CITATION = ARXIV:1711.02080;%%
\bibitem [{\citenamefont {Doneva}\ and\ \citenamefont
  {Yazadjiev}(2018)}]{Doneva:2017bvd}%
  \BibitemOpen
  \bibfield  {author} {\bibinfo {author} {\bibfnamefont {D.~D.}\ \bibnamefont
  {Doneva}}\ and\ \bibinfo {author} {\bibfnamefont {S.~S.}\ \bibnamefont
  {Yazadjiev}},\ }\href {\doibase 10.1103/PhysRevLett.120.131103} {\bibfield
  {journal} {\bibinfo  {journal} {Phys. Rev. Lett.}\ }\textbf {\bibinfo
  {volume} {120}},\ \bibinfo {pages} {131103} (\bibinfo {year} {2018})},\
  \Eprint {http://arxiv.org/abs/1711.01187} {arXiv:1711.01187 [gr-qc]}
  \BibitemShut {NoStop}%
%%CITATION = ARXIV:1711.01187;%%
\bibitem [{\citenamefont {Andreou}\ \emph {et~al.}(2019)\citenamefont
  {Andreou}, \citenamefont {Franchini}, \citenamefont {Ventagli},\ and\
  \citenamefont {Sotiriou}}]{Andreou:2019ikc}%
  \BibitemOpen
  \bibfield  {author} {\bibinfo {author} {\bibfnamefont {N.}~\bibnamefont
  {Andreou}}, \bibinfo {author} {\bibfnamefont {N.}~\bibnamefont {Franchini}},
  \bibinfo {author} {\bibfnamefont {G.}~\bibnamefont {Ventagli}}, \ and\
  \bibinfo {author} {\bibfnamefont {T.~P.}\ \bibnamefont {Sotiriou}},\ }\href
  {\doibase 10.1103/PhysRevD.99.124022} {\bibfield  {journal} {\bibinfo
  {journal} {Phys. Rev.}\ }\textbf {\bibinfo {volume} {D99}},\ \bibinfo {pages}
  {124022} (\bibinfo {year} {2019})},\ \Eprint
  {http://arxiv.org/abs/1904.06365} {arXiv:1904.06365 [gr-qc]} \BibitemShut
  {NoStop}%
%%CITATION = ARXIV:1904.06365;%%
\bibitem [{\citenamefont {Ramazanoğlu}(2017)}]{Ramazanoglu:2017xbl}%
  \BibitemOpen
  \bibfield  {author} {\bibinfo {author} {\bibfnamefont {F.~M.}\ \bibnamefont
  {Ramazanoğlu}},\ }\href {\doibase 10.1103/PhysRevD.96.064009} {\bibfield
  {journal} {\bibinfo  {journal} {Phys. Rev.}\ }\textbf {\bibinfo {volume}
  {D96}},\ \bibinfo {pages} {064009} (\bibinfo {year} {2017})},\ \Eprint
  {http://arxiv.org/abs/1706.01056} {arXiv:1706.01056 [gr-qc]} \BibitemShut
  {NoStop}%
%%CITATION = ARXIV:1706.01056;%%
\bibitem [{\citenamefont {Ramazanoğlu}(2018)}]{Ramazanoglu:2018hwk}%
  \BibitemOpen
  \bibfield  {author} {\bibinfo {author} {\bibfnamefont {F.~M.}\ \bibnamefont
  {Ramazanoğlu}},\ }\href {\doibase 10.1103/PhysRevD.98.044011} {\bibfield
  {journal} {\bibinfo  {journal} {Phys. Rev.}\ }\textbf {\bibinfo {volume}
  {D98}},\ \bibinfo {pages} {044011} (\bibinfo {year} {2018})},\ \Eprint
  {http://arxiv.org/abs/1804.00594} {arXiv:1804.00594 [gr-qc]} \BibitemShut
  {NoStop}%
%%CITATION = ARXIV:1804.00594;%%
\bibitem [{\citenamefont {Herdeiro}\ \emph {et~al.}(2018)\citenamefont
  {Herdeiro}, \citenamefont {Radu}, \citenamefont {Sanchis-Gual},\ and\
  \citenamefont {Font}}]{Herdeiro:2018wub}%
  \BibitemOpen
  \bibfield  {author} {\bibinfo {author} {\bibfnamefont {C.~A.~R.}\
  \bibnamefont {Herdeiro}}, \bibinfo {author} {\bibfnamefont {E.}~\bibnamefont
  {Radu}}, \bibinfo {author} {\bibfnamefont {N.}~\bibnamefont {Sanchis-Gual}},
  \ and\ \bibinfo {author} {\bibfnamefont {J.~A.}\ \bibnamefont {Font}},\
  }\href {\doibase 10.1103/PhysRevLett.121.101102} {\bibfield  {journal}
  {\bibinfo  {journal} {Phys. Rev. Lett.}\ }\textbf {\bibinfo {volume} {121}},\
  \bibinfo {pages} {101102} (\bibinfo {year} {2018})},\ \Eprint
  {http://arxiv.org/abs/1806.05190} {arXiv:1806.05190 [gr-qc]} \BibitemShut
  {NoStop}%
%%CITATION = ARXIV:1806.05190;%%
\bibitem [{\citenamefont {Doneva}\ \emph {et~al.}(2018)\citenamefont {Doneva},
  \citenamefont {Kiorpelidi}, \citenamefont {Nedkova}, \citenamefont
  {Papantonopoulos},\ and\ \citenamefont {Yazadjiev}}]{Doneva:2018rou}%
  \BibitemOpen
  \bibfield  {author} {\bibinfo {author} {\bibfnamefont {D.~D.}\ \bibnamefont
  {Doneva}}, \bibinfo {author} {\bibfnamefont {S.}~\bibnamefont {Kiorpelidi}},
  \bibinfo {author} {\bibfnamefont {P.~G.}\ \bibnamefont {Nedkova}}, \bibinfo
  {author} {\bibfnamefont {E.}~\bibnamefont {Papantonopoulos}}, \ and\ \bibinfo
  {author} {\bibfnamefont {S.~S.}\ \bibnamefont {Yazadjiev}},\ }\href {\doibase
  10.1103/PhysRevD.98.104056} {\bibfield  {journal} {\bibinfo  {journal} {Phys.
  Rev. D}\ }\textbf {\bibinfo {volume} {98}},\ \bibinfo {pages} {104056}
  (\bibinfo {year} {2018})},\ \Eprint {http://arxiv.org/abs/1809.00844}
  {arXiv:1809.00844 [gr-qc]} \BibitemShut {NoStop}%
\bibitem [{\citenamefont {Ventagli}\ \emph {et~al.}(2020)\citenamefont
  {Ventagli}, \citenamefont {Leh\'ebel},\ and\ \citenamefont
  {Sotiriou}}]{Ventagli:2020rnx}%
  \BibitemOpen
  \bibfield  {author} {\bibinfo {author} {\bibfnamefont {G.}~\bibnamefont
  {Ventagli}}, \bibinfo {author} {\bibfnamefont {A.}~\bibnamefont {Leh\'ebel}},
  \ and\ \bibinfo {author} {\bibfnamefont {T.~P.}\ \bibnamefont {Sotiriou}},\
  }\href {\doibase 10.1103/PhysRevD.102.024050} {\bibfield  {journal} {\bibinfo
   {journal} {Phys. Rev. D}\ }\textbf {\bibinfo {volume} {102}},\ \bibinfo
  {pages} {024050} (\bibinfo {year} {2020})},\ \Eprint
  {http://arxiv.org/abs/2006.01153} {arXiv:2006.01153 [gr-qc]} \BibitemShut
  {NoStop}%
\bibitem [{\citenamefont {Dima}\ \emph {et~al.}(2020)\citenamefont {Dima},
  \citenamefont {Barausse}, \citenamefont {Franchini},\ and\ \citenamefont
  {Sotiriou}}]{Dima:2020yac}%
  \BibitemOpen
  \bibfield  {author} {\bibinfo {author} {\bibfnamefont {A.}~\bibnamefont
  {Dima}}, \bibinfo {author} {\bibfnamefont {E.}~\bibnamefont {Barausse}},
  \bibinfo {author} {\bibfnamefont {N.}~\bibnamefont {Franchini}}, \ and\
  \bibinfo {author} {\bibfnamefont {T.~P.}\ \bibnamefont {Sotiriou}},\ }\href
  {\doibase 10.1103/PhysRevLett.125.231101} {\bibfield  {journal} {\bibinfo
  {journal} {Phys. Rev. Lett.}\ }\textbf {\bibinfo {volume} {125}},\ \bibinfo
  {pages} {231101} (\bibinfo {year} {2020})},\ \Eprint
  {http://arxiv.org/abs/2006.03095} {arXiv:2006.03095 [gr-qc]} \BibitemShut
  {NoStop}%
\bibitem [{\citenamefont {Herdeiro}\ \emph {et~al.}(2021)\citenamefont
  {Herdeiro}, \citenamefont {Radu}, \citenamefont {Silva}, \citenamefont
  {Sotiriou},\ and\ \citenamefont {Yunes}}]{Herdeiro:2020wei}%
  \BibitemOpen
  \bibfield  {author} {\bibinfo {author} {\bibfnamefont {C.~A.~R.}\
  \bibnamefont {Herdeiro}}, \bibinfo {author} {\bibfnamefont {E.}~\bibnamefont
  {Radu}}, \bibinfo {author} {\bibfnamefont {H.~O.}\ \bibnamefont {Silva}},
  \bibinfo {author} {\bibfnamefont {T.~P.}\ \bibnamefont {Sotiriou}}, \ and\
  \bibinfo {author} {\bibfnamefont {N.}~\bibnamefont {Yunes}},\ }\href
  {\doibase 10.1103/PhysRevLett.126.011103} {\bibfield  {journal} {\bibinfo
  {journal} {Phys. Rev. Lett.}\ }\textbf {\bibinfo {volume} {126}},\ \bibinfo
  {pages} {011103} (\bibinfo {year} {2021})},\ \Eprint
  {http://arxiv.org/abs/2009.03904} {arXiv:2009.03904 [gr-qc]} \BibitemShut
  {NoStop}%
\bibitem [{\citenamefont {Berti}\ \emph {et~al.}(2021)\citenamefont {Berti},
  \citenamefont {Collodel}, \citenamefont {Kleihaus},\ and\ \citenamefont
  {Kunz}}]{Berti:2020kgk}%
  \BibitemOpen
  \bibfield  {author} {\bibinfo {author} {\bibfnamefont {E.}~\bibnamefont
  {Berti}}, \bibinfo {author} {\bibfnamefont {L.~G.}\ \bibnamefont {Collodel}},
  \bibinfo {author} {\bibfnamefont {B.}~\bibnamefont {Kleihaus}}, \ and\
  \bibinfo {author} {\bibfnamefont {J.}~\bibnamefont {Kunz}},\ }\href {\doibase
  10.1103/PhysRevLett.126.011104} {\bibfield  {journal} {\bibinfo  {journal}
  {Phys. Rev. Lett.}\ }\textbf {\bibinfo {volume} {126}},\ \bibinfo {pages}
  {011104} (\bibinfo {year} {2021})},\ \Eprint
  {http://arxiv.org/abs/2009.03905} {arXiv:2009.03905 [gr-qc]} \BibitemShut
  {NoStop}%
\bibitem [{\citenamefont {Silva}\ \emph {et~al.}(2019)\citenamefont {Silva},
  \citenamefont {Macedo}, \citenamefont {Sotiriou}, \citenamefont {Gualtieri},
  \citenamefont {Sakstein},\ and\ \citenamefont {Berti}}]{Silva:2018qhn}%
  \BibitemOpen
  \bibfield  {author} {\bibinfo {author} {\bibfnamefont {H.~O.}\ \bibnamefont
  {Silva}}, \bibinfo {author} {\bibfnamefont {C.~F.~B.}\ \bibnamefont
  {Macedo}}, \bibinfo {author} {\bibfnamefont {T.~P.}\ \bibnamefont
  {Sotiriou}}, \bibinfo {author} {\bibfnamefont {L.}~\bibnamefont {Gualtieri}},
  \bibinfo {author} {\bibfnamefont {J.}~\bibnamefont {Sakstein}}, \ and\
  \bibinfo {author} {\bibfnamefont {E.}~\bibnamefont {Berti}},\ }\href
  {\doibase 10.1103/PhysRevD.99.064011} {\bibfield  {journal} {\bibinfo
  {journal} {Phys. Rev.}\ }\textbf {\bibinfo {volume} {D99}},\ \bibinfo {pages}
  {064011} (\bibinfo {year} {2019})},\ \Eprint
  {http://arxiv.org/abs/1812.05590} {arXiv:1812.05590 [gr-qc]} \BibitemShut
  {NoStop}%
%%CITATION = ARXIV:1812.05590;%%
\bibitem [{\citenamefont {Macedo}\ \emph {et~al.}(2019)\citenamefont {Macedo},
  \citenamefont {Sakstein}, \citenamefont {Berti}, \citenamefont {Gualtieri},
  \citenamefont {Silva},\ and\ \citenamefont {Sotiriou}}]{Macedo:2019sem}%
  \BibitemOpen
  \bibfield  {author} {\bibinfo {author} {\bibfnamefont {C.~F.~B.}\
  \bibnamefont {Macedo}}, \bibinfo {author} {\bibfnamefont {J.}~\bibnamefont
  {Sakstein}}, \bibinfo {author} {\bibfnamefont {E.}~\bibnamefont {Berti}},
  \bibinfo {author} {\bibfnamefont {L.}~\bibnamefont {Gualtieri}}, \bibinfo
  {author} {\bibfnamefont {H.~O.}\ \bibnamefont {Silva}}, \ and\ \bibinfo
  {author} {\bibfnamefont {T.~P.}\ \bibnamefont {Sotiriou}},\ }\href@noop {} {\
   (\bibinfo {year} {2019})},\ \Eprint {http://arxiv.org/abs/1903.06784}
  {arXiv:1903.06784 [gr-qc]} \BibitemShut {NoStop}%
%%CITATION = ARXIV:1903.06784;%%
\bibitem [{\citenamefont {Antoniou}\ \emph {et~al.}(2021)\citenamefont
  {Antoniou}, \citenamefont {Bordin},\ and\ \citenamefont
  {Sotiriou}}]{Antoniou:2020nax}%
  \BibitemOpen
  \bibfield  {author} {\bibinfo {author} {\bibfnamefont {G.}~\bibnamefont
  {Antoniou}}, \bibinfo {author} {\bibfnamefont {L.}~\bibnamefont {Bordin}}, \
  and\ \bibinfo {author} {\bibfnamefont {T.~P.}\ \bibnamefont {Sotiriou}},\
  }\href {\doibase 10.1103/PhysRevD.103.024012} {\bibfield  {journal} {\bibinfo
   {journal} {Phys. Rev. D}\ }\textbf {\bibinfo {volume} {103}},\ \bibinfo
  {pages} {024012} (\bibinfo {year} {2021})},\ \Eprint
  {http://arxiv.org/abs/2004.14985} {arXiv:2004.14985 [gr-qc]} \BibitemShut
  {NoStop}%
\bibitem [{\citenamefont {Kanti}\ \emph {et~al.}(1996)\citenamefont {Kanti},
  \citenamefont {Mavromatos}, \citenamefont {Rizos}, \citenamefont {Tamvakis},\
  and\ \citenamefont {Winstanley}}]{Kanti:1995vq}%
  \BibitemOpen
  \bibfield  {author} {\bibinfo {author} {\bibfnamefont {P.}~\bibnamefont
  {Kanti}}, \bibinfo {author} {\bibfnamefont {N.~E.}\ \bibnamefont
  {Mavromatos}}, \bibinfo {author} {\bibfnamefont {J.}~\bibnamefont {Rizos}},
  \bibinfo {author} {\bibfnamefont {K.}~\bibnamefont {Tamvakis}}, \ and\
  \bibinfo {author} {\bibfnamefont {E.}~\bibnamefont {Winstanley}},\ }\href
  {\doibase 10.1103/PhysRevD.54.5049} {\bibfield  {journal} {\bibinfo
  {journal} {Phys. Rev. D}\ }\textbf {\bibinfo {volume} {54}},\ \bibinfo
  {pages} {5049} (\bibinfo {year} {1996})},\ \Eprint
  {http://arxiv.org/abs/hep-th/9511071} {arXiv:hep-th/9511071} \BibitemShut
  {NoStop}%
\bibitem [{\citenamefont {Sotiriou}\ and\ \citenamefont
  {Zhou}(2014)}]{Sotiriou:2014pfa}%
  \BibitemOpen
  \bibfield  {author} {\bibinfo {author} {\bibfnamefont {T.~P.}\ \bibnamefont
  {Sotiriou}}\ and\ \bibinfo {author} {\bibfnamefont {S.-Y.}\ \bibnamefont
  {Zhou}},\ }\href {\doibase 10.1103/PhysRevD.90.124063} {\bibfield  {journal}
  {\bibinfo  {journal} {Phys. Rev.}\ }\textbf {\bibinfo {volume} {D90}},\
  \bibinfo {pages} {124063} (\bibinfo {year} {2014})},\ \Eprint
  {http://arxiv.org/abs/1408.1698} {arXiv:1408.1698 [gr-qc]} \BibitemShut
  {NoStop}%
%%CITATION = ARXIV:1408.1698;%%
\bibitem [{\citenamefont {Blázquez-Salcedo}\ \emph {et~al.}(2018)\citenamefont
  {Blázquez-Salcedo}, \citenamefont {Doneva}, \citenamefont {Kunz},\ and\
  \citenamefont {Yazadjiev}}]{Blazquez-Salcedo:2018jnn}%
  \BibitemOpen
  \bibfield  {author} {\bibinfo {author} {\bibfnamefont {J.~L.}\ \bibnamefont
  {Blázquez-Salcedo}}, \bibinfo {author} {\bibfnamefont {D.~D.}\ \bibnamefont
  {Doneva}}, \bibinfo {author} {\bibfnamefont {J.}~\bibnamefont {Kunz}}, \ and\
  \bibinfo {author} {\bibfnamefont {S.~S.}\ \bibnamefont {Yazadjiev}},\ }\href
  {\doibase 10.1103/PhysRevD.98.084011} {\bibfield  {journal} {\bibinfo
  {journal} {Phys. Rev.}\ }\textbf {\bibinfo {volume} {D98}},\ \bibinfo {pages}
  {084011} (\bibinfo {year} {2018})},\ \Eprint
  {http://arxiv.org/abs/1805.05755} {arXiv:1805.05755 [gr-qc]} \BibitemShut
  {NoStop}%
%%CITATION = ARXIV:1805.05755;%%
\bibitem [{\citenamefont {Maselli}\ \emph {et~al.}(2020)\citenamefont
  {Maselli}, \citenamefont {Franchini}, \citenamefont {Gualtieri},\ and\
  \citenamefont {Sotiriou}}]{Maselli:2020zgv}%
  \BibitemOpen
  \bibfield  {author} {\bibinfo {author} {\bibfnamefont {A.}~\bibnamefont
  {Maselli}}, \bibinfo {author} {\bibfnamefont {N.}~\bibnamefont {Franchini}},
  \bibinfo {author} {\bibfnamefont {L.}~\bibnamefont {Gualtieri}}, \ and\
  \bibinfo {author} {\bibfnamefont {T.~P.}\ \bibnamefont {Sotiriou}},\ }\href
  {\doibase 10.1103/PhysRevLett.125.141101} {\bibfield  {journal} {\bibinfo
  {journal} {Phys. Rev. Lett.}\ }\textbf {\bibinfo {volume} {125}},\ \bibinfo
  {pages} {141101} (\bibinfo {year} {2020})},\ \Eprint
  {http://arxiv.org/abs/2004.11895} {arXiv:2004.11895 [gr-qc]} \BibitemShut
  {NoStop}%
\bibitem [{\citenamefont {Ripley}\ and\ \citenamefont
  {Pretorius}(2020)}]{Ripley:2020vpk}%
  \BibitemOpen
  \bibfield  {author} {\bibinfo {author} {\bibfnamefont {J.~L.}\ \bibnamefont
  {Ripley}}\ and\ \bibinfo {author} {\bibfnamefont {F.}~\bibnamefont
  {Pretorius}},\ }\href {\doibase 10.1088/1361-6382/ab9bbb} {\bibfield
  {journal} {\bibinfo  {journal} {Class. Quant. Grav.}\ }\textbf {\bibinfo
  {volume} {37}},\ \bibinfo {pages} {155003} (\bibinfo {year} {2020})},\
  \Eprint {http://arxiv.org/abs/2005.05417} {arXiv:2005.05417 [gr-qc]}
  \BibitemShut {NoStop}%
\end{thebibliography}%

\end{document}